\documentclass[12pt]{article}
\usepackage[a4paper, total={7in, 10in}]{geometry}
\usepackage[parfill]{parskip}
\usepackage{physics, tensor, float, subcaption}
\usepackage{graphicx}
\graphicspath{ {Plots/} }
\usepackage{jhep-mod}
\usepackage{bm}
\usepackage{soul}
\usepackage{amssymb,amsmath,amsthm}
\usepackage{mathrsfs}
\usepackage[utf8]{inputenc}
\usepackage{enumerate}
\usepackage{bigints}
\usepackage{xcolor}
\usepackage{appendix}
\usepackage{graphicx}
\usepackage{float}
\usepackage{tikz}
\usepackage{setspace}
\usepackage{cancel}
\usepackage{doi}
\definecolor{purple}{rgb}{1,0,1}
\definecolor{lime}{HTML}{A6CE39} 

\definecolor{lime}{HTML}{A6CE39}
\newcommand{\orcidicon}{%
	\begin{tikzpicture}
	\draw[lime, fill=lime] (0,0) 
		circle [radius=0.16] 
		node[white] {{\fontfamily{qag}\selectfont \tiny ID}};
	\draw[white, fill=white] (-0.0625,0.095) 
		circle [radius=0.007];
	\end{tikzpicture}
	\hspace{-5mm}
}
\newcommand\orcidJosh{{\href{https://orcid.org/0000-0003-1200-7261}{\orcidicon}}}
\newcommand\orcidMatt{{\href{https://orcid.org/0000-0003-1088-6485}{\orcidicon}}}

\renewcommand{\O}{\mathcal{O}}
\newcommand{\C}{\mathcal{C}}

\begin{document}


\title{\vspace{-25pt}\huge{
Physically motivated ansatz\\ for the Kerr spacetime
}}


\author{
\Large
Joshua Baines\!\orcidJosh\!, 
{\sf  and} Matt Visser\!\orcidMatt\!}
\affiliation{
School of Mathematics and Statistics, Victoria University of Wellington, 
\\
\null\qquad PO Box 600, Wellington 6140, New Zealand.}
\emailAdd{joshua.baines@sms.vuw.ac.nz}
\emailAdd{matt.visser@sms.vuw.ac.nz}

\abstract{
\vspace{1em}

Despite some 60 years of work on the subject of the Kerr rotating black hole there is as yet no widely accepted physically based and pedagogically viable ansatz suitable for deriving the Kerr solution without significant computational effort. (Typically involving computer-aided symbolic algebra.) Perhaps the closest one gets in this regard is the Newman--Janis trick; a trick which requires several physically unmotivated choices in order to work. Herein we shall try to make some progress on this issue by using a \emph{non-ortho-normal} tetrad based on oblate spheroidal coordinates to absorb as much of the messy angular dependence as possible, leaving one to deal with a relatively simple angle-independent tetrad-component metric. 
That is, we shall write $g_{ab} = g_{AB} \; e^A{}_a\; e^B{}_b$ seeking to keep both the tetrad-component metric $g_{AB}$ and the \emph{non-ortho-normal} co-tetrad $e^A{}_a$ relatively simple but non-trivial. We shall see that it is possible to put all the mass dependence into $g_{AB}$, while the \emph{non-ortho-normal} co-tetrad $e^A{}_a$ can be chosen to be a mass-independent representation of flat Minkowski space in oblate spheroidal coordinates: $(g_\mathrm{Minkowski})_{ab} = \eta_{AB} \; e^A{}_a\; e^B{}_b$. This procedure separates out, to the greatest extent possible, the mass dependence from the rotational dependence, and makes the Kerr solution perhaps a little less mysterious. 

\bigskip

\bigskip
\noindent
{\sc Date:} 19 July 2022; 25 July 2022; \LaTeX-ed \today

\bigskip
\noindent{\sc Keywords}: \\
Kerr spacetime; Kerr--Newman spacetime; oblate spheroidal coordinates; \\
non-ortho-normal tetrads; Newman--Janis trick.

\bigskip
\noindent{\sc PhySH:} 
Gravitation
}

\maketitle
\def\tr{{\mathrm{tr}}}
\def\diag{{\mathrm{diag}}}
\def\cof{{\mathrm{cof}}}
\def\pdet{{\mathrm{pdet}}}
\def\d{{\mathrm{d}}}
\def\L{{\mathcal{L}}}
\parindent0pt
\parskip7pt
\renewcommand{\C}{\mathcal{C}}
\newcommand{\arctanh}{\tanh^{-1}}
\newcommand{\n}{\nabla}

\clearpage
\null
\vspace{-75pt} 
\section{Introduction}

The Kerr solution was discovered in 1963~\cite{Kerr,Kerr-Texas}, and quickly became a mainstay of general relativity, though it took the wider astrophysics community somewhat longer to appreciate its full significance. Understanding how the Kerr solution was first discovered is tricky~\cite{Kerr-history}, and even to this day no really clean pedagogical first-principles derivation exists. 
Typically one tells the students: ``Here is the answer, feed it into your favourite computer algebra system [{\sf Maple}, {\sf Mathematica}, {\sf Wolfram Alpha}, \emph{whatever}] and check that the Ricci tensor is zero.''
Interest in the Kerr spacetime is both intense and ongoing, with many review articles~\cite{kerr-intro, Teukolsky:2014, Adamo:2014,  Hehl:2014, Bambi:2011, Reynolds:2013, Johannsen:2015,  Bambi:2017, Reynolds:2019}, at least two dedicated books~\cite{kerr-book, kerr-book-2}, and many textbook discussions~\cite{Weinberg, MTW, Adler-Bazin-Schiffer, Wald, D'Inverno, Hartle, Carroll,  Hobson,  Poisson, Padmanabhan}.

The closest one has to a pedagogical first-principles derivation of the Kerr spacetime is via the Newman--Janis trick~\cite{Newman-Janis,NJ-wiki}, developed in 1965,  which was immediately used in then deriving the electro-magnetically charged Kerr--Newman spacetime~\cite{Kerr-Newman}. 
Despite many valiant efforts~\cite{Newman:1977, Giampieri:1990, Drake:1997, Drake:1998, Viaggiu:2006, Hansen:2013, Ferraro:2013, Keane:2014, Erbin:2014, Erbin:2016, Rajan:2016-NJ} it is still fair to say that no fully convincing explanation of why the Newman--Janis trick works has been forthcoming.\footnote{A somewhat different ansatz, based on rather strong assumptions regarding the geodesics, has been explored in references~\cite{Dadhich:2013,Dadhich:2022}.} 

\enlargethispage{49pt}
Herein we shall try a different approach: 
\begin{itemize}
\item First, since we know that for a Newtonian rotating fluid body the Maclaurin spheroid is a good first approximation~\cite{Spheroid, Maclaurin, Chandrasekhar, Poisson-Will, Lyttleton}, and that this is an example of an oblate spheroid in flat 3-space, one strongly suspects that oblate spheroidal coordinates might be useful when it comes to investigating rotating black holes. (And for that matter, other rotating bodies in general relativity.)
\item
Second, we know that the tetrad formalism is extremely useful~\cite{Rajan:2016-global}, both in purely classical general relativity and especially when working with elementary particles with spin.
\item
Third, in a rather different context, the use of \emph{non-ortho-normal} tetrads has recently proved to be extremely useful
\cite{Visser:2021}.
\end{itemize}
These observations \emph{suggest} that it might be useful to write the spacetime metric in the form
\begin{equation}
g_{ab} = g_{AB} \;\; e^A{}_a\;\; e^B{}_b,
\end{equation}
where we shall seek to push all of the mass dependence into the tetrad-component metric $g_{AB}$, while pushing as much as possible of the rotational aspects of the problem into the \emph{non-ortho-normal} co-tetrad $e^A{}_a$.
Specifically we shall show that we can choose the \emph{non-ortho-normal} co-tetrad to represent flat Minkowski space in oblate spheroidal coordinates
\begin{equation}
(g_\mathrm{Minkowski})_{ab} = 
\eta_{AB} \;\; e^A{}_a\;\; e^B{}_b;
\qquad 
\eta_{AB} = \mathrm{diag}\{-1,1,1,1\}.
\end{equation}
This procedure separates out, to the greatest extent possible, the mass dependence from the rotational dependence, and makes the Kerr solution perhaps a little less mysterious. 

\clearpage
\section{Preliminaries}
\enlargethispage{40pt}
In any spacetime manifold one can always (at least locally) set up a flat metric $(g_\mathrm{flat})_{ab}$ and from that flat metric extract a (non-unique) co-tetrad 
\begin{equation}
(g_\mathrm{flat})_{ab} = 
\eta_{AB} \;\: e^A{}_a\;\: e^B{}_b;
\qquad 
\eta_{AB} = \mathrm{diag}\{-1,1,1,1\}.
\end{equation}
Given such a co-tetrad, one can construct the associated tetrad $e_A{}^a$, (which is just the matrix inverse of the co-tetrad) and  then for any arbitrary (non-flat) metric $g_{ab}$ one can always write: 
\begin{equation}
g_{ab}=g_{AB}\;\; e^A{}_a\; \;e^B{}_b \,;
\qquad
g_{AB} = g_{ab} \;\; e_A{}^a\;\; e_B{}^b. 
\end{equation}
We wish to derive the Kerr solution by finding a suitable flat-space tetrad, then make a natural and simple ansatz for $g_{AB}$, and check that $g_{ab}$ satisfies the vacuum Einstein equations $R_{ab}=0$. 

\paragraph{Example:} Consider Schwarzschild spacetime. 
Order the coordinates as $\{t,r,\theta,\phi\}$. Take the co-tetrad for flat space written in spherical polar coordinates to be:
\begin{equation}
e^A{}_a=\left[\begin{array}{cccc}
1 & 0 & 0 & 0 \\
0 & 1 & 0 & 0 \\
0 & 0 & r & 0 \\
0 & 0 & 0 & r\sin\theta \\
\end{array}\right]\,.
\end{equation}
Then, given the symmetries of the spacetime, a natural and simple ansatz for the tetrad-component metric $g_{AB}$ would be 
\begin{equation}
g_{AB}=\left[\begin{array}{cccc}
-f(r) & 0 & 0 & 0 \\
0 & \frac{1}{f(r)} &\;\; 0 & 0 \\
0 & 0 & 1 & 0 \\
0 & 0 & 0 & 1 \\
\end{array}\right]\,.
\end{equation}
In this situation the Einstein equations yield
\begin{equation}
f(r)=1-\frac{2m}{r}\,.
\end{equation}
Similarly, for the Reissner--Nordstr\"om spacetime one simply has
\begin{equation}
f(r)=1-\frac{2m}{r} + {Q^2\over r^2}\,.
\end{equation}
We shall now seek to do something similar for the Kerr spacetime.

\clearpage
\section{Oblate spheroidal coordinates} 
The Cartesian metric for flat spacetime can be rewritten in terms of oblate spheroidal coordinates by defining
\begin{equation}
\begin{split}
x & =\sqrt{r^2+a^2}\;\sin\theta\;\cos\phi \,;\\
y & =\sqrt{r^2+a^2}\;\sin\theta\;\sin\phi \,;\\
z & =r\cos\theta\,.
\end{split}
\end{equation}
Then, ordering the coordinates as $\{t,r,\theta,\phi\}$, the metric is 
\begin{equation}
\label{E:oblate-minkowski}
g_{ab}=\left[\begin{array}{cccc}
-1 & 0 & 0 & 0 \\
0 & \frac{r^2+a^2\cos^2\theta}{r^2+a^2} & 0 & 0 \\
0 & 0 & r^2+a^2\cos^2\theta & 0 \\
0 & 0 & 0 & (r^2+a^2)\sin^2\theta \\
\end{array}\right]\,.
\end{equation}
Setting $A\in\{0,1,2,3\}$, an obvious (but naive) co-tetrad for this metric is 
\begin{equation}
\label{ob_tet_flat}
(e_\mathrm{naive})^A{}_a=\left[\begin{array}{cccc}
1 & 0 & 0 & 0 \\
0 & \sqrt{\frac{r^2+a^2\cos^2\theta}{r^2+a^2}} & 0 & 0 \\
0 & 0 & \sqrt{r^2+a^2\cos^2\theta} & 0 \\
0 & 0 & 0 & \sqrt{r^2+a^2}\sin\theta \\
\end{array}\right]\,.
\end{equation}
However, trying to use this co-tetrad would not be ideal for deriving the Kerr spacetime since the Kerr spacetime is \emph{stationary} not \emph{static}, meaning that the Kerr metric must have non-zero, off-diagonal components. 
\enlargethispage{20pt}

We could introduce non-diagonal components in our ansatz for  the tetrad-component metric $g_{AB}$, however this vastly complicates the computations. In order to simplify the derivation, we will instead find a non-diagonal co-tetrad $e^A{}_a$ which will allow us to make $g_{AB}$ diagonal. More specifically, we will make an ansatz of the form  
\begin{equation}
\label{Kerr_Guess}
g_{AB}=\left[\begin{array}{cccc}
-f(r) & 0 & 0 & 0 \\
0 & \frac{1}{f(r)} & 0 & 0 \\
0 & 0 & 1 & 0 \\
0 & 0 & 0 & 1 \\
\end{array}\right]\,,
\end{equation}
as we did for the Schwarzschild and Reissner--Nordstr\"om spacetimes.

\clearpage
\section{An improved co-tetrad }

Note that there exist an infinite number of co-tetrads for any given spacetime metric, related via local Lorentz transformations.  (That is,  if $e^A{}_a$ is a co-tetrad, then so is $L^A{}_B\; e^B{}_a$, where $L^A{}_B$ is a tangent-space Lorentz transformation). We wish to transform the naive tetrad given in equation \eqref{ob_tet_flat} via a Lorentz transformation into a more useful form. 

Since we are using an ansatz for $g_{AB}$  of the form given in equation \eqref{Kerr_Guess}, we wish the $e^0{}_t$ component of our new tetrad to be the reciprocal of the $e^1{}_r$ component. 
Furthermore, we will ask that the $g_{t\phi}$ component  be the only non-zero off-diagonal component of our final spacetime metric $g_{ab}$. This then constrains the relevant local Lorentz transformation to be of the form
\begin{equation} 
L^A{}_B=\left[\begin{array}{cccc}
\sqrt{\frac{r^2+a^2}{r^2+a^2\cos^2\theta}} &\; 0 & 0 & \;L^0{}_3 \\
0 & 1 & 0 & 0 \\
0 & 0 & 1 & 0 \\
L^3{}_0 & 0 & 0 & L^3{}_3 \\
\end{array}\right]\,.
\end{equation} 
However, since $L^A{}_B$ is a Lorentz transformation, it must satisfy
\begin{equation} 
L^C{}_A\;\; \eta_{CD}\;\; L^D{}_B = \eta_{AB}\,.
\end{equation} 
This tightly constrains the components of $L^A{}_B$; in fact this requirement can be used to solve for the remaining 3 components. They are given by
\begin{equation} 
\begin{split}
L^3{}_3 & = L^1{}_1 = \sqrt{\frac{r^2+a^2}{r^2+a^2\cos^2\theta}} \,;\\
L^3{}_0 & = L^0{}_3= -\frac{a\sin\theta}{\sqrt{r^2+a^2\cos^2\theta}} \,.
\end{split}
\end{equation} 
Here $a$ can be either positive or negative depending on the sense of rotation.

Hence, explicitly, we have
\begin{equation} 
L^A{}_B=\left[\begin{array}{cccc}
\sqrt{\frac{r^2+a^2}{r^2+a^2\cos^2\theta}} & 0 & 0 & -\frac{a\sin\theta}{\sqrt{r^2+a^2\cos^2\theta}} \\
0 & 1 & 0 & 0 \\
0 & 0 & 1 & 0 \\
-\frac{a\sin\theta}{\sqrt{r^2+a^2\cos^2\theta}} & 0 & 0 & \sqrt{\frac{r^2+a^2}{r^2+a^2\cos^2\theta}} \\
\end{array}\right]\,.
\end{equation} 
Note that $\det( L^A{}_B)=1$ and that this local Lorentz transformation corresponds to the velocity $\beta = 
{a\sin\theta\over\sqrt{r^2+a^2}} \in (-1,+1)$.
 
\clearpage
\null
\vspace{-75pt} 
Our new improved co-tetrad is now given by
\enlargethispage{20pt}
\begin{equation} 
\begin{split}
{e}^A{}_a & = L^A{}_B\;\; (e_\mathrm{naive})^B{}_a \\[7pt]
& =\left[\begin{array}{cccc}
\sqrt{\frac{r^2+a^2}{r^2+a^2\cos^2\theta}} & 0 & 0 & -\frac{\sqrt{r^2+a^2}a\sin^2\theta}{\sqrt{r^2+a^2\cos^2\theta}} \\
0 & \sqrt{\frac{r^2+a^2\cos^2\theta}{r^2+a^2}} & 0 & 0 \\
0 & 0 & \sqrt{r^2+a^2\cos^2\theta} & 0 \\
-\frac{a\sin\theta}{\sqrt{r^2+a^2\cos^2\theta}} & 0 & 0 & \frac{(r^2+a^2)\sin\theta}{\sqrt{r^2+a^2\cos^2\theta}} \\
\end{array}\right]\,.
\end{split}
\end{equation} 
Using the ansatz for $g_{AB}$, as given in equation \eqref{Kerr_Guess}, we now find
\begin{equation} 
\begin{split}
g_{ab} & =g_{AB}\;\; {e}^A{}_a\;\; {e}^B{}_b \\[7pt]
& =\left[\begin{array}{cccc}
\frac{-f(r)(r^2+a^2)+a^2\sin^2\theta}{r^2+a^2\cos^2\theta} & 0 & 0 & g_{t\phi} \\
0 & \frac{r^2+a^2\cos^2\theta}{f(r)(r^2+a^2)} & 0 & 0 \\
0 & 0 & r^2+a^2\cos^2\theta & 0 \\
g_{t\phi} & 0 & 0 & g_{\phi\phi} \\
\end{array}\right]\,.
\end{split}
\end{equation} 
Here
\begin{equation} 
g_{t\phi}=\frac{(r^2+a^2)\;a\sin^2\theta\;(f(r)-1)}{r^2+a^2\cos^2\theta}\,,
\end{equation} 
and
\begin{equation} 
g_{\phi\phi}=\frac{(r^2+a^2)\;\sin^2\theta\;(r^2+a^2-f(r)a^2\sin^2\theta)}{r^2+a^2\cos^2\theta}\,.
\end{equation} 

\section{Final step: Einstein equations }

We now apply the Einstein equations to the ansatz developed above.
The vacuum Einstein equations then give a system of (partial) differential equations for the metric components $g_{ab}$. 
Explicitly finding the set of PDEs is still best done with a computer algebra system, but one now has a well-defined and relatively simple problem to solve, and the PDEs reduce to ODEs for the function $f(r)$. 
The simplest of these ODEs is 
\begin{equation} 
R_{\theta\theta}=\frac{df(r)}{dr}(r^3+ra^2)+f(r)(r^2-a^2)-r^2+a^2=0\,.
\end{equation} 
This is a first-order linear ODE which has the solution
\begin{equation} 
f(r)=1-\frac{2mr}{r^2+a^2}\,.
\end{equation} 
This  finally results in the fully explicit metric
\begin{equation}  
\label{Kerr}
g_{ab}=\left[\begin{array}{cccc}
-\left(1-\frac{2mr}{\rho^2}\right) & \;0 & 0 & \;-\frac{2mra\sin^2\theta}{\rho^2} \\
0 & \frac{\rho^2}{\Delta} & 0 & 0 \\
0 & 0 & \rho^2 & 0 \\
-\frac{2mra\sin^2\theta}{\rho^2} & 0 & 0 & \Sigma\sin^2\theta \\
\end{array}\right]\,.
\end{equation} 
Here (as usual) we have $\rho=\sqrt{r^2+a^2\cos^2\theta}$,\; while $\Delta=r^2+a^2-2mr$, and in turn $\Sigma=r^2+a^2+2mra^2\sin^2\theta/\rho^2$. Notice that equation \eqref{Kerr} is just the Kerr metric written in the usual Boyer--Lindquist coordinates, which hence concludes the derivation. 

\section{Summary} 

The Kerr metric (and the Minkowski metric) can be related to the mass-independent co-tetrad 
\begin{equation} 
{e}^A{}_a  
 =\left[\begin{array}{cccc}
\sqrt{\frac{r^2+a^2}{r^2+a^2\cos^2\theta}} & 0 & 0 & -\frac{\sqrt{r^2+a^2}a\sin^2\theta}{\sqrt{r^2+a^2\cos^2\theta}} \\
0 & \sqrt{\frac{r^2+a^2\cos^2\theta}{r^2+a^2}} & 0 & 0 \\
0 & 0 & \sqrt{r^2+a^2\cos^2\theta} & 0 \\
-\frac{a\sin\theta}{\sqrt{r^2+a^2\cos^2\theta}} & 0 & 0 & \frac{(r^2+a^2)\sin\theta}{\sqrt{r^2+a^2\cos^2\theta}} \\
\end{array}\right]\,.
\end{equation} 
by the very simple relations
\begin{equation}
(g_\mathrm{Kerr})_{ab} = g_{AB} \;\; e^A{}_a\;\; e^B{}_b;
\qquad
(g_\mathrm{Minkowski})_{ab} = \eta_{AB} \;\; e^A{}_a\;\; e^B{}_b;
\end{equation}
where the tetrad-component metric is particularly simple
\begin{equation}
\label{E:basic}
g_{AB}=\left[\begin{array}{cccc}
-f(r) & 0 & 0 & 0 \\
0 & \frac{1}{f(r)} & 0 & 0 \\
0 & 0 & 1 & 0 \\
0 & 0 & 0 & 1 \\
\end{array}\right]\,; \qquad
f(r)=1-\frac{2mr}{r^2+a^2}\,.
\end{equation}
This manifestly has the appropriate limit as $a\to 0$  and 
cleanly separates out the angular and radial behaviour. Furthermore the only change needed to accomodate the electromagnetically charged Kerr--Newman solution is to replace $2mr \to 2mr -Q^2$ and so to set
\begin{equation}
f(r)=1-\frac{2mr- Q^2}{r^2+a^2}\,.
\end{equation}
We have been relatively slow and careful in developing and presenting the analysis, trying to provide physical motivations for our choices at each step of the process. Of course, once you see the answer, the reason it works is obvious in hindsight --- simply take the usual Boyer--Lindquist \emph{ortho-normal} co-tetrad for Kerr, set $m\to 0$, and then set $\eta_{AB}\to g_{AB}$ to compensate. 
Given this, can we now generalize the ansatz to deal with other coordinate representations~\cite{Baines:unit-lapse, Baines:Darboux}   of the Kerr spacetime? 
(Or its slow-rotation Lense--Thirring~\cite{Lense-Thirring, Pfister, PGLT1, PGLT2, PGLT3, PGLT4} approximation?)

\clearpage

\section{Extensions of the basic ansatz} 

We now develop several extensions and generalizations of the basic ansatz (\ref{E:basic}) presented above.

\subsection{Eddington--Finkelstein (Kerr--Schild) form}

Take
\begin{equation}
\label{E:EF-KS}
g_{AB}=\left[\begin{array}{cccc}
-1+\Phi & \Phi & 0 & 0 \\
\Phi & 1+\Phi & 0 & 0 \\
0 & 0 & 1 & 0 \\
0 & 0 & 0 & 1 \\
\end{array}\right]\,; \qquad
\Phi = {2mr\over r^2+a^2}\,.
\end{equation}
Keep exactly the same non-ortho-normal co-tetrad $e^A{}_a$ as above. 
Then the metric $g_{ab} = g_{AB} \; e^A{}_a\; e^B{}_b$ is still Ricci flat --- so it is the Kerr solution in disguise. 
This modified ansatz was inspired by inspecting and generalizing the Eddington--Finkelstein (Kerr--Schild) form of Schwarzschild. 
Defining $\ell_A = (1,1,0,0)$ we note that
\begin{equation}
g_{AB} = \eta_{AB} + \Phi \;\ell_A\; \ell_B
\end{equation}
which is of Kerr--Schild form. Contracting with the co-tetrad and defining $\ell_a = \ell_A \; e^A{}_a$ we have
\begin{equation}
g_{ab} = (g_\mathrm{Minkowski})_{ab} + \Phi \;\ell_a \; \ell_b,
\end{equation}
where the Minkowski space metric is written in oblate spheroidal coordinates as per (\ref{E:oblate-minkowski}) and 
\begin{equation}
\ell_a = \left(\sqrt{r^2+a^2\over r^2+a^2 \cos^2\theta} ,
\sqrt{r^2+a^2\cos^2\theta\over r^2+a^2} , 0,
- a \sin^2\theta \sqrt{r^2+a^2\over r^2+a^2 \cos^2\theta}
 \right).
\end{equation}
Since $\ell_a$ is easily checked to be a null vector, this is manifestly seen to be Kerr spacetime in Kerr--Schild form~\cite{kerr-intro, Kerr-history, kerr-book}.

\subsection{Quasi-Painlev\'e--Gullstrand form}

Take
\begin{equation}
\label{E:qPG}
g_{AB}=\left[\begin{array}{cccc}
-1+\Phi &\; \sqrt{\Phi} & 0 & 0 \\
\sqrt{\Phi} & 1 & \; 0 & 0 \\
0 & 0 & 1 & 0 \\
0 & 0 & 0 & 1 \\
\end{array}\right]\,; \qquad
\Phi = {2mr\over r^2+a^2}\,.
\end{equation}
Keep exactly the same non-ortho-normal co-tetrad $e^A{}_a$ as above.
Then the metric $g_{ab} = g_{AB} \; e^A{}_a\; e^B{}_b$ is still Ricci flat --- so it is the Kerr solution in disguise. 

\clearpage
This modified ansatz was \emph{inspired} by looking at and generalizing the Painlev\'e-Gullstrand  form of Schwarzschild; though it was not \emph{derived} therefrom --- more on this point later.
The tetrad metric $g_{AB}$ is of Painlev\'e--Gullstrand  form,
but the coordinate metric $g_{ab}$ is not, (and, in view of the analysis by Valiente-Kroon~\cite{Valiente-Kroon:2003,Valiente-Kroon:2004}, cannot possibly be),  of  Painlev\'e--Gullstrand  form.

It is convenient to introduce two vectors, $T_A=(1,0,0,0)$ and $S_A=(0,1,0,0)$, since then 
\begin{equation}
g_{AB} = \eta_{AB} + \Phi \;T_A\; T_B 
+ \sqrt{\Phi} \;(T_A \;S_B + S_A \; T_B).
\end{equation}
We can furthermore factorize this as follows
\begin{equation}
g_{AB} = \eta_{CD} \; 
\left(\delta^C{}_A + \sqrt{\Phi} \;S^C \;T_A \right) \;
\left(\delta^D{}_B + \sqrt{\Phi} \;S^D \;T_B \right),
\end{equation}
implying the existence of a factorizable \emph{ortho-normal} co-tetrad
\begin{equation}
(e_{ortho})^A{}_ a =  
\left(\delta^A{}_B + \sqrt{\Phi} \; S^A \; T_B \right) \; e^B{}_ a.
\end{equation}
Note that all the mass-dependence is concentrated in $\Phi$, whereas all the angular dependence is still concentrated in the usual mass-independent 
\emph{non-ortho-normal} tetrad $e^B{}_ a$.

Let us see what happens in the coordinate basis: Setting 
\begin{equation}
T_a = T_A\; e^A{}_a = \left(\sqrt{r^2+a^2\over r^2+a^2 \cos^2\theta} ,
0 , 0,
- a \sin^2\theta \sqrt{r^2+a^2\over r^2+a^2 \cos^2\theta}
\right),
\end{equation}
and 
\begin{equation}
S_a = S_A\; e^A{}_a = \left(0,  \sqrt{r^2+a^2\cos^2\theta\over r^2+a^2} ,0,0
\right),
\end{equation}
we see
\begin{equation}
\label{E:QPG-full}
g_{ab} = (g_\mathrm{Minkowski})_{ab} + \Phi \;T_a \; T_b
+ 
\sqrt{\Phi} \;(T_a \;S_b + S_a \; T_b),
\end{equation}
where the Minkowski space metric is written in oblate spheroidal coordinates as per (\ref{E:oblate-minkowski}).
The vectors $T_a$ and $S_a$ are orthonormal timelike and spacelike vectors with respect to both the Minkowski metric
(\ref{E:oblate-minkowski}) and the full metric (\ref{E:QPG-full}). In the language of the Hamilton--Lisle ``river model''~\cite{Hamilton:2004} these are easily identified as what they call the ``twist'' and ``flow'' vectors, and so this quasi-Painlev\'e--Gullstrand  version of the Kerr metric is equivalent to the Doran form~\cite{Doran:1999} of the Kerr metric.
This is the closest we can get to putting the Kerr metric into Painlev\'e--Gullstrand form --- partial success at the tetrad level, but failure at the coordinate level. 
Finally, we observe that explicit computation reveals that $g^{tt}=-1$, so this version of the metric is definitely unit lapse~\cite{Baines:unit-lapse}. 

\subsection{1-free-function form}

Let $h(r)$ be an arbitrary differentiable function and take
\begin{equation}
\label{E:1-free}
g_{AB}=\left[\begin{array}{cccc}
-f(r)& -f(r) h(r) & 0 & 0 \\
\; -f(r) h(r) & \;{1\over f(r)} - f(r) h(r)^2 \;& 0 & 0 \\
0 & 0 & 1 & 0 \\
0 & 0 & 0 & 1 \\
\end{array}\right]\,; \qquad
f(r) = 1-{2mr\over r^2+a^2}\,.
\end{equation}
Keep exactly the same non-ortho-normal co-tetrad $e^A{}_a$ as above. 
Then the metric $g_{ab} = g_{AB} \; e^A{}_a\; e^B{}_b$ is still Ricci flat --- so it is the Kerr solution in disguise. 

This ansatz was \emph{inspired} by looking at and generalizing the Boyer--Lindquist, Kerr--Schild, and quasi-Painlev\'e--Gullstrand  forms of Kerr discussed above; not \emph{derived} therefrom. With hindsight, one strongly suspects an underlying coordinate transformation is responsible for this behaviour. Indeed after a little ``reverse engineering'' one is lead to consider the not particularly obvious coordinate transformation
\begin{equation}
\label{E:specific}
t \to t + \int h(r) \; \d r; \qquad \phi \to \phi + \int {a\, h(r)\over r^2+a^2} \; \d r. 
\end{equation}
Writing the new coordinates as $\bar x^a$ the relevant Jacobi matrix is
\begin{equation}
J^a{}_b = (\bar x^a)_{,b} 
= {\partial\bar x^a\over \partial x^b}  =
\left[ \begin{array}{cccc}
1 & h(r) &0&0\\0&1&0&0\\0&0&1&0\\
0 &{a\, h(r)\over r^2+a^2} &0&1
\end{array} \right].
\end{equation}
Going to the tetrad basis an easy computation yields
\begin{equation}
J^A{}_B = J^a{}_b \;\; e_a{}^A\;\;  e^b{}_B =
\left[ \begin{array}{cccc}
1 & h(r) &0&0\\0&1&0&0\\0&0&1&0\\
0 &0&0&1
\end{array} \right].
\end{equation}
But then it is easy to check that
\begin{eqnarray}
\left[ \begin{array}{cccc}
1 & h(r) &0&0\\0&1&0&0\\0&0&1&0\\
0 &0&0&1
\end{array} \right]^T
\left[\begin{array}{cccc}
-f(r) & 0 & 0 & 0 \\
0 & \frac{1}{f(r)} & 0 & 0 \\
0 & 0 & 1 & 0 \\
0 & 0 & 0 & 1 \\
\end{array}\right]
\left[ \begin{array}{cccc}
1 & h(r) &0&0\\0&1&0&0\\0&0&1&0\\
0 &0&0&1
\end{array} \right] 
\nonumber\\
=
\left[\begin{array}{cccc}
-f(r)& -f(r) h(r) & 0 & 0 \\
\; -f(r) h(r) & \;{1\over f(r)} - f(r) h(r)^2 \;& 0 & 0 \\
0 & 0 & 1 & 0 \\
0 & 0 & 0 & 1 \\
\end{array}\right].
\end{eqnarray}
That is, the 1-free-function ansatz (\ref{E:1-free}) can be obtained from the basic ansatz (\ref{E:basic}) by the very specific coordinate transformation (\ref{E:specific}); with the  
specific coordinate transformation being carefully ``reverse engineered'' to do minimal violence to the original basic ansatz.

\subsection{Lense--Thirring limit}

Consider the slow rotation limit $a\to 0$, explicitly keeping the first two terms, while keeping $h(r)$ arbitrary, then
\begin{equation} 
{e}^A{}_a 
 =\left[\begin{array}{cccc}
 1 & 0 & \;0 & -a\sin^2\theta \\
0 & 1 & 0 & 0 \\
0 & 0 & r & 0 \\
-\frac{a\sin\theta}{r} & 0 & 0 & r\sin\theta \\
\end{array}\right] + \O(a^3)\,.
\end{equation} 
and 
\begin{equation}
g_{AB}=\left[\begin{array}{cccc}
-f(r)& -f(r) h(r) & 0 & 0 \\
\; -f(r) h(r) & \;{1\over f(r)} - f(r) h(r)^2 \;& 0 & 0 \\
0 & 0 & 1 & 0 \\
0 & 0 & 0 & 1 \\
\end{array}\right]\,; \qquad
f(r)=1-\frac{2m}{r} + {2m a^2\over r^3}+ \O(a^4)\,.
\end{equation}
Thus our 1-free-function ansatz (\ref{E:1-free}) leads to an entire class of tetrad metrics $g_{AB}$,
(and implicitly, the corresponding coordinate basis metrics $g_{ab}$), that are appropriate for describing the exterior spacetime of slowly rotating objects. 
This Lense--Thirring slow rotation limit~\cite{Lense-Thirring,Pfister}, in its various incarnations~\cite{PGLT1,PGLT2,PGLT3,PGLT4} is of significant importance in observational astrophysics.

\clearpage
\subsection{Summary}

In this section we have seen how our basic ansatz (\ref{E:basic}), which we originally developed for physically motivating and then deriving the Kerr solution with a minimum of fuss, can be extended and modified to deal with other coordinate representations of the Kerr metric --- such as the Eddington--Finkelstein (Kerr--Schild) coordinates (\ref{E:EF-KS}), the quasi-Painlev\'e--Gullstrand (Doran) coordinates (\ref{E:qPG}), and an entire 1-free-function class of coordinate systems (\ref{E:1-free}) that still respect most of the fundamental symmetries of the original ansatz.

\section{Discussion} 

We have physically motivated an ansatz for the Kerr spacetime metric, partially based on Newtonian physics, (the fact that Maclaurin's oblate spheroids already became of interest for rotating bodies some 280 years ago), and partially based on the fact that tetrad methods are known to be useful in general relativity.
Specifically, the key step is to write the coordinate metric as $g_{ab} = g_{AB}\; e^A{}_a\; e^B{}_b$, while allowing the use of \emph{non-ortho-normal} tetrads. 

We have seen that doing so permits one to force all of the non-trivial angular dependence into a mass-independent co-tetrad $e^A{}_a$ that is compatible with flat spacetime in oblate spheroidal coordinates, while forcing all of the mass-dependence (and none of the angular dependence) into the tetrad-basis metric $g_{AB}$. This clean separation between angular dependence and mass dependence greatly simplifies the computational complexity of the problem. 
We expect these ideas to have further applications and implications. 

\bigskip
\hrule\hrule\hrule

\section*{Acknowledgements}

JB was supported by a Victoria University of Wellington PhD Doctoral Scholarship
and was also indirectly supported by the Marsden Fund, 
via a grant administered by the Royal Society of New Zealand.
\\
MV was directly supported by the Marsden Fund, 
via a grant administered by the Royal Society of New Zealand.

\bigskip
\hrule\hrule\hrule

\clearpage

\end{document}